\documentclass[review,sort&compress]{elsarticle}
\usepackage{latexsym,amssymb,graphicx,epstopdf,epsf,epsfig}
\usepackage{amsmath,graphicx,caption,subcaption,float}
\usepackage{geometry,natbib,pifont}
\journal{Digital Signal Processing}

\newtheorem{thm}{Theorem}
\newtheorem{lem}{Lemma}
\newtheorem{cor}{Corollary}
\newtheorem{prop}{Proposition}
\newproof{pot1}{Proof of Theorem \ref{thm1}}
\newproof{pot2}{Proof of Theorem \ref{thm3}}
\newproof{pot3}{Proof of Theorem \ref{thm3}}
\newproof{pol1}{Proof of Lemma \ref{lem1}}
\newproof{pol2}{Proof of Lemma \ref{lem2}}
\newproof{poc1}{Outline of the Proof of Corollary \ref{cor1}}
\newproof{poc2}{Outline of the Proof of Corollary \ref{cor2}}
\newproof{pop1}{Proof of Proposition \ref{prop1}}
\newdefinition{rmk}{Remark}
\newproof{pf}{Proof}

\DeclareMathOperator*{\argmin}{arg\,min}

\renewcommand{\vec}[1]{\mbox{$\mathbf{#1}$}}

\newcommand{\norm}[1]{\left|\left|#1\right|\right|}
\newcommand{\defi}{\triangleq}

\newcommand{\tr}{\mathrm{Tr}}

\newcommand{\col}{\mathrm{vec}}
\newcommand{\nn}{\nonumber}

\newcommand{\C}{\mathbb{C}}
\newcommand{\R}{{\cal R}}

\newcommand{\vH}{\vec{H}}
\newcommand{\vx}{\vec{x}}
\newcommand{\vy}{\vec{y}}
\newcommand{\dH}{\Delta\vH}
\newcommand{\dy}{\Delta\vy}
\newcommand{\tH}{\tilde{\vec{H}}}
\newcommand{\ty}{\tilde{\vec{y}}}
\newcommand{\tP}{\tilde{\vec{P}}}
\newcommand{\vP}{\vec{P}}
\newcommand{\vD}{\vec{D}}
\newcommand{\vd}{\vec{d}}
\newcommand{\dvh}{\Delta\vec{h}}
\newcommand{\vX}{\vec{X}}
\newcommand{\vI}{\vec{I}}
\newcommand{\vb}{\vec{b}}
\newcommand{\vw}{\vec{w}}
\newcommand{\vh}{\vec{h}}
\newcommand{\vQ}{\vec{Q}}

\newcommand{\vG}{\vec{G}}
\newcommand{\vN}{\vec{N}}
\newcommand{\vZ}{\vec{Z}}
\newcommand{\vu}{\vec{u}}
\newcommand{\vF}{\vec{F}}
\newcommand{\vW}{\vec{W}}

\newcommand{\va}{\mbox{\boldmath$\alpha$}}
\newcommand{\vbet}{\mbox{\boldmath$\beta$}}

\begin{document}

\begin{frontmatter}

\title{Robust Least Squares Methods Under Bounded Data Uncertainties}

\author[1]{N. Denizcan Vanli\corref{cor}}
\ead{vanli@ee.bilkent.edu.tr}
\author[2]{Mehmet A. Donmez}
\ead{donmez2@illinois.edu}
\author[1]{Suleyman S. Kozat}
\ead{kozat@bilkent.edu.tr}
\cortext[cor]{Corresponding Author}
\address[1]{Department of Electrical and Electronics Engineering, Bilkent University, Ankara, Tel: +90-312-290-2336.}
\address[2]{Department of Electrical and Computer Engineering, University of Illinois at Urbana-Champaign, Illinois.}

\begin{abstract}
  We study the problem of estimating an unknown deterministic signal that is observed through an unknown deterministic data matrix under additive noise. In particular, we present a minimax optimization framework to the least squares problems, where the estimator has imperfect data matrix and output vector information. We define the performance of an estimator relative to the performance of the optimal least squares (LS) estimator tuned to the underlying unknown data matrix and output vector, which is defined as the regret of the estimator. We then introduce an efficient robust LS estimation approach that minimizes this regret for the worst possible data matrix and output vector, where we refrain from any structural assumptions on the data. We demonstrate that minimizing this worst-case regret can be cast as a semi-definite programming (SDP) problem. We then consider the regularized and structured LS problems and present novel robust estimation methods by demonstrating that these problems can also be cast as SDP problems. We illustrate the merits of the proposed algorithms with respect to the well-known alternatives in the literature through our simulations.
\end{abstract}
\end{frontmatter}

\section{Introduction}
In this paper we investigate the estimation of an unknown deterministic signal that is observed through a deterministic data matrix under additive noise \cite{kuruoglu1,donmez,dsp1,dsp2,dsp3,dsp4,kuruoglu2,kalantarova,Ghaoui97,Kailath:book,yonina,sayedchandra1,sayedchandra2,sayedchandra3}. Although the data matrix and the output vector are not exactly known, estimates for both of them as well as uncertainty bounds on the estimates are given \cite{pilanci10,sayed02,donmez,kalantarova,yonina2,yonina4}. When the model parameters are not known exactly, a popular method to estimate the desired signal is to use the robust LS method \cite{Ghaoui97}, since the performances of the classical LS estimators degrade significantly when the perturbations on the data matrix and the output vector are relatively high \cite{Ghaoui97,sayed02,pilanci10,kuruoglu3,schubert,huber1}.

A prevalent approach to find robust solutions to such estimation problems is the robust LS method \cite{Ghaoui97, sayed02,kalantarova}, in which the uncertainties in the data matrix and the output vector are incorporated into optimization framework via a minimax residual formulation. Another well-known approach to compensate for errors in the data matrix and the output vector is the total least squares method (TLS) \cite{pilanci10}, which may yield undesirable results since it employs a conservative approach due to data de-regularization. Furthermore, the data matrix usually has a known special structure, such as Toeplitz and Hankel, in many linear regression problems \cite{Ghaoui97, pilanci10} and the performance of the estimators based on minimax approaches are shown to improve
when such a prior knowledge on data matrix structure is integrated into the problem formulation \cite{Ghaoui97, pilanci10}.

Although the robust LS methods are able to minimize the LS error for the worst-case perturbations, they usually provide unsatisfactory results on the average \cite{pilanci10,chandra1,chandra2,chandra3} due to their conservative nature. In order to counterbalance this conservative nature of the robust LS methods \cite{Ghaoui97}, we propose a novel robust LS approach that minimizes a worst case ``regret'' that is defined as the difference between the squared residual error and the smallest attainable squared residual error with an LS estimator. By this regret formulation, we seek a linear estimator whose performance is as close as possible to that of the optimal estimator for all possible perturbations on the data matrix and the output vector. Our main goal in proposing the minimax regret formulation is to provide a trade-off between the robust LS methods tuned to the worst possible data parameters (under the uncertainty bounds) and the optimal LS estimator tuned to the underlying unknown model parameters. Furthermore, after studying the data estimation problems in the presence of bounded data uncertainties, we extend the regret formulation to the regularized LS problem, where the regret is defined as the difference between the cost of the regularized LS algorithm \cite{Kailath:book, sayed02}, and the smallest attainable cost with a linear regularized LS estimator. Furthermore, we extend our discussions to scenarios involving both structured and unstructured data. Under these frameworks, we provide the solutions for the proposed regret based minimax LS and the regret based minimax regularized LS approaches in semi-definite programming (SDP) forms. We emphasize that SDP problems can be efficiently solved even for real-time applications \cite{boyd}.

Minimax regret approaches have been presented in signal processing literature to alleviate the pessimistic nature of the worst case optimization methods \cite{ElMe04, ElTaNe04, eldar_mrr, Kozat:09TSP,donmez,kalantarova}. However, we emphasize that the methods proposed in this paper extensively differ from \cite{Ghaoui97, sayed02, ElMe04, ElTaNe04,Kozat:09TSP,donmez}. Note that the optimization frameworks investigated here are different than \cite{Ghaoui97, sayed02}, where the regret terms are directly adjoined in the cost functions. Although a similar regret notion is used in \cite{ElMe04,ElTaNe04,Kozat:09TSP,donmez}, the cost function as well as the constraints on uncertainties in the data matrix and the output vector are substantially different in this paper. Moreover, unlike this paper, in \cite{donmez}, the problem is described for the channel equalization scenario, where the authors rely on the statistical assumptions. Furthermore, we note that the uncertainty is in the statistics of the transmitted signal in \cite{ElMe04}. In \cite{ElTaNe04} and \cite{Kozat:09TSP}, the uncertainty is in the transmitted signal and the channel parameters, respectively. Unlike these relevant works, in this paper, the uncertainty is both on the data matrix and the output vector. Furthermore, the solutions to the LS problems presented in this paper cannot be obtained from \cite{ElMe04, ElTaNe04, Kozat:09TSP,donmez,kalantarova}, since the cost functions are different in our optimization formulations. While in \cite{kalantarova}, the authors have considered a similar framework, the results of this paper builds upon them and provide a complete solution to the regret based robust LS estimation methods unlike \cite{kalantarova}. We emphasize that perturbation bounds on the data matrix and the output vector heavily depend on the estimation algorithms employed to obtain them. Since our methods are formulated for given perturbation bounds, different estimation algorithms can be readily incorporated into our framework with the corresponding perturbation bounds \cite{sayed02}.

In this paper, we first present a novel robust LS approach in which we seek to find the transmitted signal by minimizing the worst case regret, i.e., the worst case difference between the residual error of the LS estimator and the residual error with the optimal LS estimator. In this sense, our aim is to introduce a trade off between the performance of the robust LS methods and the tuned LS estimator (LS estimator that is tuned to the unknown data matrix and the output vector). We next propose a minimax regret method for the regularized LS problem. Finally, we introduce a structured robust LS method in which the data matrix has a special structure such as Toeplitz and Hankel. We demonstrate that the proposed robust methods can be cast as SDP problems. In our simulations, we observe that these approaches provide better performance compared to the robust methods that are optimized with respect to the worst-case residual error \cite{Ghaoui97, sayed02}, the tuned LS and the tuned regularized LS estimators (tuned to the estimates of the data matrix and the output vector), respectively.

Our main contributions in this paper are as follows. {\em i)} We introduce a novel and efficient robust LS estimation method in which we find the transmitted signal by minimizing the worst-case regret, i.e., the worst-case difference
between the residual error of the LS estimator and the residual error of the optimal LS estimator tuned to the underlying model. In this sense, we present a robust estimation method that achieves a tradeoff between the robust LS estimation methods and the direct LS estimation method tuned to the estimates of the data matrix and output vector. {\em ii)} We next propose a minimax regret formulation for the regularized LS estimation problem. {\em iii)} We then introduce a structured robust LS estimation method in which the data matrix is known to have a special structure such as Toeplitz or Hankel. {\em iv)} We demonstrate that the robust estimation methods we propose can be cast as SDP problems, hence our methods can be efficiently applied for real-time \cite{boyd}. {\em iv)} In our simulations, we observe that our approaches provide better performance compared to the robust methods that are optimized with respect to the worst-case residual error \cite{est1, est2}, and the conventional methods that directly solve the estimation problem using the perturbed data.

The organization of the paper is as follows. An overview to the problem is provided in Section \ref{sec:system}. In Section \ref{sec:urls}, we first introduce the LS estimation method based on our regret formulation, and then present the regularized LS estimation approach in Section \ref{sec:urrls}. We then consider the structured LS approach in Section \ref{sec:srls} and provide the explicit SDP formulations for all problems. The numerical examples are demonstrated in Section \ref{sec:numer}. Finally, the paper concludes with certain remarks in Section \ref{sec:conc}.

\section{System Overview} \label{sec:system}
\subsection{Notation}
In this paper, all vectors are column vectors and represented by boldface lowercase letters. Matrices are represented by boldface uppercase letters. For a matrix $\vH$, $\vH^H$ is the conjugate transpose, $\norm{\vH}$ is the spectral norm, $\vH^+$ is the pseudo-inverse, $\vH > 0$ represents a positive definite matrix and $\vH \geq 0$ represents a positive semi-definite matrix. For a square matrix $\vH$, $\tr(\vH)$ is the trace. Naturally, for a vector $\vx$, $\norm{\vx} = \sqrt{\vx^H \vx}$ is the $\ell^2$-norm. Here, $\vec{0}$ denotes a vector or matrix with all zero elements and the dimensions can be understood from the context. Similarly, $\vI$ represents the appropriate sized identity matrix. The operator $\col(\cdot)$ is the vectorization operator, i.e., it stacks the columns of a matrix of dimension $m\times n$ into a $mn \times 1$ column vector. Finally, the operator $\otimes$ is the Kronecker product \cite{graham}.

\subsection{Problem Description}
We investigate the problem of estimating an unknown deterministic vector $\vx \in \mathbb{C}^{n}$ which is observed through a deterministic data matrix. However, instead of the actual data matrix and the output vector, their estimates $\vH \in \C^{m\times n}$ and $\vy \in \C^m$ and uncertainty bounds on these estimates are provided. In this sense, our aim is to find a solution to the following data estimation problem
\[
\vy \approx \vH \vx,
\]
such that
\[
\vy+\dy = (\vH+\dH)\vx,
\]
for deterministic perturbations $\dH \in \C^{m \times n}$, $\dy \in \C^m$. Although these perturbations are unknown, a bound on each perturbation is provided, i.e.,
\[
\norm{\dH} \leq \delta_H \text{ and } \norm{\dy} \leq \delta_Y,
\]
where $\delta_H, \delta_Y \geq 0$. In this sense, we refrain from any assumptions on the data matrix and the output vector, yet consider that the estimates $\vH$ and $\vy$ are at least accurate to ``some degree'' but their actual values under these uncertainties are completely unknown to the estimator.

Even in the presence of these uncertainties, the symbol vector $\vx$ can be naively estimated by simply substituting the estimates $\vH$ and $\vy$ into the LS estimator \cite{sayedbook}. For the LS estimator we have
\[
\hat{\vx} = \vH^ + \vy,
\]
where $\vH^+$ is the pseudo-inverse of $\vH$ \cite{graham}. However, this approach yields unsatisfactory results, when the errors in the estimates of the data matrix and the output vector are relatively high \cite{est1, est2, yonina1, yonina2, yonina3}. A common approach to find a robust solution is to employ a worst-case residual minimization \cite{est1}
\[
\hat{\vx} = \argmin_{\vx \in \C^n} \max_{\norm{\dH} \leq \delta_H, \norm{\dy} \leq \delta_Y} {\norm{(\vy+\dy)-(\vH + \dH)\vx}}^2,
\]
where $\vx$ is chosen to minimize the worst-case residual error in the uncertainty region. However, since the solution is found with respect to the worst possible data matrix and output vector in the uncertainty regions, it may be highly conservative \cite{tls, yonina1, yonina3}.

Here, we propose a novel LS estimation approach that provides a tradeoff between performance and robustness in order to mitigate the conservative nature of the worst-case residual minimization approach as well as to preserve robustness \cite{yonina1, yonina3}. The regret for not using the optimal LS estimator is defined as the difference between the residual error with an estimate of the input vector and the residual error with the optimal LS estimator, i.e.,
\begin{equation}\label{eq:LSreg0}
  \R(\vx; \dH,\dy) \defi {\norm{(\vy+\dy)-(\vH + \dH)\vx}}^2 - \min_{\vw \in \C^n}{\norm{(\vy+\dy)-(\vH +\dH)\vw}}^2.
\end{equation}
By making such a regret definition, we force our estimator not to construct the symbol vector according to the worst possible scenario considering that it may be too conservative. Instead, we define the regret of any estimator by the difference in the estimation performances of that estimator and the ``smartest'' estimator knowing both data matrix and output vector in hindsight, so that we achieve a tradeoff between robustness and estimation performance.

We emphasize that the regret defined in \eqref{eq:LSreg0} is completely different than the regret formulation introduced in \cite{yonina1, yonina3}. In \eqref{eq:LSreg0}, the uncertainty is on the data matrix where the desired data vector $\vx$ is completely unknown, unlike \cite{yonina1, yonina3}. We emphasize that we use the residual error ${\norm{(\vy+\dy)-(\vH + \dH)\vx}}^2$ instead of the estimation error $\norm{\hat{\vx} - \vx}$ since the estimation error directly depends on the vector $\vx$ and cannot be used in the regret formulation since $\vx$ is assumed to be unknown in the presence of data uncertainties. Moreover, in our formulation, the estimate $\hat{\vx}$ is not constrained to be linear unlike \cite{yonina1, yonina3} since our regret formulation is well-defined without any limitations on the estimated $\hat{\vx}$.

In the next sections, the proposed approaches to the robust LS estimation problems are provided. We first introduce the regret based unstructured LS estimation method. We next present the unstructured regularized LS estimation approach in which the worst-case regret is optimized. Finally, we investigate the structured LS estimation approach.

\section{Unstructured Robust Least Squares Estimation} \label{sec:urls}
In this section, we provide a novel robust unstructured LS estimator based on a certain minimax criterion. We consider the most generic estimation problem
\begin{equation}\label{eq:prudef1}
  \min_{\vx \in \C^n} \max_{\norm{\dH} \leq \delta_H,\norm{\dy} \leq \delta_Y} \R(\vx; \dH,\dy),
\end{equation}
where $\R(\vx; \dH,\dy)$ is defined as in \eqref{eq:LSreg0}. Now considering the second term in \eqref{eq:LSreg0}, we define $\tH \defi \vH + \dH$, $\ty \defi \vy+\dy$, where $\tH$ is a full rank matrix, and denote the estimation performance of the optimal LS estimator for some given $\tH$ and $\ty$ by
\[
f(\tH, \ty) \defi \min_{\vw \in \C^n}{\norm{\ty - \tH \vw}}^2.
\]
Since we consider an unconstrained minimization over $\vw$, we have \cite{sayedbook}
\begin{align}
  \vw^* & \defi \argmin_{\vw \in \C^n}{\norm{\ty - \tH \vw}}^2 \nn\\
        & = \tH^+ \ty,
\end{align}
as the optimal data vector minimizing the residual error. Then we have
\begin{align}
  f\left(\tH, \ty \right) & = \norm{\ty - \tH \vw^*}^2 \nn\\ %\norm{\vP \, \vy}^2
              & = (\ty - \tH \vw^*)^H (\ty - \tH \vw^*) \nn\\
              & = \ty^H (\ty - \tH \vw^*) \nn\\
              & = \ty^H \tP \ty, \nn
\end{align}
where the third line follows from $\tH^H \tH \vw^* = \tH^H \ty$ \cite{sayedbook} and $\tP \defi \vI-\tH\tH^+$ is the projection matrix of the space perpendicular to the range space of $\tH$. If we use the Taylor series expansion based on Wirtinger calculus \cite{graham} for $f\left(\tH, \ty \right)$ around $\tH = \vH$ and $\ty = \vy$, then
\begin{equation}\label{eq:1stTaylorApp}
  f\left(\tH, \ty \right) = f(\vH,\vy) + 2\operatorname{Re} \left\{ \tr\left( \nabla f(\tH, \ty)\big\vert_{\tH=\vH, \ty = \vy}^H \big[\dH \;\; \dy\big] \right) \right\} + O\left( \norm{\left[\dH \hspace{0.1in} \dy \right]}^2 \right).
\end{equation}
Note that the first order Taylor approximation is introduced in order to obtain a tractable solution. Clearly, the effect of using this approximation vanishes as $\norm{[\dH \;\; \dy ]}$ decreases and for distortions with larger $\norm{[\dH \;\; \dy ]}$, one can easily use higher order approximations instead. However, we observe through our simulations that even for relatively large perturbations, a satisfactory performance is obtained using this approximation.

We now introduce the following lemma in order to obtain the first order Taylor approximation in \eqref{eq:1stTaylorApp} in a closed form.

\begin{lem}\label{lem1}
Let $\tH =\vH+\dH$ be a full rank matrix and $\ty=\vy+\dy$, where $\tH \in \C^{m \times n}$ and $\ty \in \C^m$. Then defining $f\left(\tH, \ty \right) \triangleq \ty^H\tP\ty$, where $\tP \triangleq \vI-\tH\tH^+$, we have
\[
\dfrac{\partial f\left(\tH, \ty \right)}{\partial \tH}\bigg\vert_{\tH=\vH, \ty=\vy} = - \vP \vy \left( \vH^+ \vy \right)^H,
\]
and
\[
\dfrac{\partial f\left(\tH, \ty \right)}{\partial \ty}\bigg\vert_{\tH=\vH, \ty=\vy} = \vP \vy,
\]
where $\vP \triangleq \vI-\vH\vH^+$
\end{lem}

\begin{pol1}
Since $\tH$ is full rank and $m \geq n$, the pseudo-inverse of $\tH$ is found by \cite{graham}
\[
\tH^+ \defi (\tH^H\tH)^{-1}\tH^H.
\]
Hence, we have \cite{graham}
\begin{align}\label{eq:D_definition}
  \vD & = \dfrac{\partial}{\partial \tH} \left( \ty^H \ty - \ty^H \tH (\tH^H\tH)^{-1}\tH^H \ty \right) \bigg\vert_{\tH=\vH, \ty=\vy} \nn\\
      & = \vH(\vH^H\vH)^{-1}\vH^H\vy\vy^H\vH(\vH^H\vH)^{-1} \nn\\
      & \hspace{0.5cm} - \vy\vy^H \vH(\vH^H \vH)^{-1} \nn\\
      & = \vH \vH^+ \vy \left( \vH^+ \vy \right)^H - \vy \left( \vH^+ \vy \right)^H \nn\\
      & = - \vP \vy \left( \vH^+ \vy \right)^H,
\end{align}
and
\begin{align}\label{eq:b_definition}
  \vb & = \dfrac{\partial}{\partial \ty} \left( \ty^H \ty - \ty^H \tH (\tH^H\tH)^{-1}\tH^H \ty \right) \bigg\vert_{\tH=\vH, \ty=\vy} \nn\\
      & = \vP \vy,
\end{align}
where the last line of the equality follows since $\vH \vH^+$ is a symmetric matrix according to the definition of the pseudo-inverse operation. This concludes the proof of Lemma 1. \hfill $\square$
\end{pol1}

Now turning our attention back to \eqref{eq:1stTaylorApp}, we denote
\[
\vD \defi \dfrac{\partial f\left(\tH, \ty \right)}{\partial \tH}\bigg\vert_{\tH=\vH, \ty=\vy},
\]
and
\[
\vb \defi \dfrac{\partial f\left(\tH, \ty \right)}{\partial \ty}\bigg\vert_{\tH=\vH, \ty=\vy},
\]
where we emphasize that the closed form definitions of $\vD$ and $\vb$ can be obtained from Lemma 1. We then approximate \eqref{eq:1stTaylorApp} and obtain the first order Taylor approximation as follows
\begin{align}\label{eq:thmReg}
  f\left(\tH, \ty \right) & \approx f(\vH, \vy) + 2\operatorname{Re}\left\{ \mathrm{Tr} \left( \left[\vD \;\; \vb \right]^H \left[\dH \;\; \dy \right] \right) \right\} \nn\\
              & = \kappa + 2\operatorname{Re}\left\{ \left(\col(\vD)^H \col(\dH) + \vb^H \dy \right) \right\} \nn\\
              & = \kappa + \vd^H \dvh + \dvh^H \vd + \vb^H \dy + \dy^H \vb,
\end{align}
where $\kappa \defi f(\vH,\vy)$, $\vd \defi \col(\vD)$, and $\dvh \defi \col(\dH)$. Hence we can approximate the regret in \eqref{eq:LSreg0} as follows
\begin{equation}\label{eq:regret_final}
  \R(\vx; \dH,\dy) \approx {\norm{\ty - \tH \vx}}^2 - \left( \kappa + \vd^H \dvh + \dvh^H \vd + \vb^H \dy + \dy^H \vb \right).
\end{equation}

In the following theorem, we illustrate how the optimization (or equivalently estimation) problem in \eqref{eq:regret_final} can be put in an SDP form.

\begin{thm}\label{thm1}
Let $\vH \in \C^{m\times n}$ and $\vy \in \C^{m}$ be the estimates of the data matrix and the output vector, respectively, both having deterministic additive perturbations $\dH \leq \delta_H$ and $\dy \leq \delta_Y$, respectively, i.e., $\tH=\vH+\dH$ and $\ty=\vy+\dy$, where $\tH$ is the full rank data matrix, $\ty$ is the output vector, and $m \geq n$. Then the problem
\begin{equation}\label{eq:prudefthm1}
  \min_{\vx \in \C^n} \max_{\norm{\dH} \leq \delta_H,\norm{\dy} \leq \delta_Y} \R(\vx; \dH,\dy),
\end{equation}
where $\R(\vx; \dH,\dy)$ is defined as in \eqref{eq:regret_final}, is equivalent to solving the following SDP problem
\begin{gather}
  \min \gamma \nn\\
  \mbox{subject to} \nn\\
  \tau_1 \geq 0, \tau_2 \geq 0 \text{, and} \nn\\
  \begin{bmatrix}
  \gamma + \kappa - \tau_1 - \tau_2 & (\vy-\vH\vx)^H & \delta_Y \vb^H  & \delta_H\vd^H \\
  \vy-\vH\vx                        & \vI            & -\delta_Y\vI    & \delta_H\vX \\
  \delta_Y \vb                      & -\delta_Y\vI   & \tau_1 \vI      & \vec{0} \\
  \delta_H\vd                       & \delta_H\vX^H  & \vec{0}         & \tau_2 \vI
  \end{bmatrix} \geq 0,
\end{gather}
where $\vX$ is the $m\times mn$ matrix defined as $\vX \defi \vx^H \otimes \vI$.
\end{thm}

The proof of Theorem 1 is provided in \ref{app:pot1}.

\begin{rmk}
In the proof of Theorem 1, we use Proposition 1 that relies on the lossless {\em S}-procedure. However, {\em S}-procedure is lossless with two constraints when the corresponding two quadratic (Hermitian) forms on the complex linear space \cite{s1}. However, classical {\em S}-procedure for quadratic forms is, in general, lossy with two constraints in the real case \cite{s2}. Hence, Theorem 1 cannot be extended for real linear space.
\end{rmk}

Now we can consider two important corollaries of Theorem 1. First, a special case of Theorem 1 in which the uncertainty is only in the data matrix. We emphasize that the perturbation errors only in the data matrix are also common in a wide range of real life applications \cite{sayedbook}. Here, we can define the regret as follows
\begin{equation}\label{eq:regretH}
\R(\vx; \dH) \defi {\norm{\vy - \tH \vx}}^2 - \min_{\vw \in \C^n}{\norm{\vy - \tH \vw}}^2,
\end{equation}
and similar to the previous case, we calculate the optimal estimation performance under a given uncertainty bound
\begin{align}
  f\left(\tH\right) & \defi \min_{\vw \in \C^n}{\norm{\vy - \tH \vw}}^2 \nn\\
         & \approx \kappa + 2\operatorname{Re} \left\{ \tr\left( \nabla f(\tH, \vy)\big\vert_{\tH=\vH}^H \;\; \dH \right) \right\} \nn\\
         & = \kappa + 2\operatorname{Re}\left\{ \col(\vD^H) \col(\dH) \right\} \nn\\
         & = \kappa + \vd^H \dvh + \dvh^H \vd. \nn
\end{align}
Hence we approximate the regret in \eqref{eq:regretH} as follows
\begin{equation}\label{eq:regret_final_H}
  \R(\vx; \dH) \approx {\norm{\vy - \tH \vx}}^2 - \left( \kappa + \vd^H \dvh + \dvh^H \vd \right).
\end{equation}

\begin{cor}\label{cor1}
Let $\vH \in \C^{m\times n}$ and $\vy \in \C^{m}$ be the estimates of the data matrix and the output vector, respectively, where $m \geq n$. Suppose there is a bounded uncertainty on the full rank data matrix $\tH$, i.e., $\tH=\vH+\dH$, $\norm{\dH} \leq \delta_H$. Then the problem
\begin{equation}\label{prudefcor1}
\min_{\vx \in \C^n} \max_{\norm{\dH} \leq \delta_H} \R(\vx; \dH),
\end{equation}
where $\R(\vx; \dH)$ is defined as in \eqref{eq:regret_final_H}, is equivalent to solving the following SDP problem
\begin{gather}
\min \gamma \nn\\
\mbox{subject to} \nn\\
\tau \geq 0 \text{ and }
\begin{bmatrix}
\gamma + \kappa -\tau  & (\vy-\vH\vx)^H  & \delta_H\vd \\
\vy-\vH\vx             & \vI             & \delta_H\vX\\
\delta_H\vd            & \delta_H\vX^H   & \tau \vI
\end{bmatrix}\geq 0.
\end{gather}
\end{cor}

\begin{poc1}
The proof of Corollary 1 can be explicitly derived from the proof of Theorem 1 by simply setting $\delta_Y = 0$ and $\tau_1 = 0$, hence is omitted. \hfill $\square$
\end{poc1}

Second, we consider another special case of Theorem 1 in which the uncertainty is only in the output vector. We emphasize that similar to the previous case, this one is also a common case in a wide range of real-life applications \cite{sayedbook}, and studied under a similar framework in \cite{yonina1}. Here, we can define the regret as follows
\begin{equation}\label{eq:regrety}
  \R(\vx; \dy) \defi {\norm{\ty - \vH \vx}}^2 - \min_{\vw \in \C^n}{\norm{\ty - \vH \vw}}^2,
\end{equation}
and similar to the previous case, we calculate the optimal also performance under a given uncertainty bound
\begin{align}
  f(\ty) & \defi \min_{\vw \in \C^n}{\norm{\ty - \vH \vw}}^2 \nn\\
         & \approx \kappa + 2\operatorname{Re} \left\{ \tr\left( \nabla f(\vH, \ty)\big\vert_{\ty=\vy}^H \;\; \dy \right) \right\} \nn\\
         & = \kappa + 2\operatorname{Re}\left\{ \vb^H \dy \right\} \nn\\
         & = \kappa + \vb^H \dy + \dy^H \vb. \nn
\end{align}
Hence we approximate the regret in \eqref{eq:regrety} as follows
\begin{equation}\label{eq:regret_final_y}
\R(\vx; \dy) \approx {\norm{\ty - \vH \vx}}^2 - \left( \kappa + \vb^H \dy + \dy^H \vb \right).
\end{equation}

\begin{cor}\label{cor2}
Let $\vH \in \C^{m\times n}$ and $\vy \in \C^{m}$ be the estimates of the data matrix and the output vector, respectively, where $m \geq n$. Suppose there is a bounded uncertainty on the output vector $\ty$, i.e., $\ty=\vy+\dy$, $\norm{\dy} \leq \delta_Y$. Then the problem
\begin{equation}
\min_{\vx \in \C^n} \max_{\norm{\dy} \leq \delta_Y} \R(\vx; \dy),
\end{equation}
where $\R(\vx; \dy)$ is defined as in \eqref{eq:regret_final_y}, is equivalent to solving the following SDP problem
\begin{gather}
\min \gamma \nn\\
\mbox{subject to} \nn\\
\tau \geq 0 \text{ and }
\begin{bmatrix}
\gamma + \kappa -\tau  & (\vy-\vH\vx)^H  & \delta_Y\vb^H \\
\vy-\vH\vx             & \vI             & -\delta_Y\vI \\
\delta_Y\vb            & -\delta_Y\vI    & \tau \vI
\end{bmatrix} \geq 0.
\end{gather}
\end{cor}

\begin{poc2}
The proof of Corollary 2 can be explicitly derived from the proof of Theorem 1 by simply setting $\delta_H = 0$ and $\tau_2 = 0$, hence is omitted. \hfill $\square$
\end{poc2}

\begin{rmk}
Corollaries 1 and 2 follows from the proof of Theorem 1, which relies on the lossless {\em S}-procedure. Under the frameworks presented in the Corollaries 1 and 2, one can safely extend the same conclusions for the real case also, since {\em S}-procedure is lossless for quadratic forms with one constraint both in complex and real spaces \cite{huang1,huang2}.
\end{rmk}

\section{Unstructured Robust Regularized Least Squares Estimation} \label{sec:urrls}
In this section, we introduce a worst-case regret optimization approach to solve the regularized LS estimation problem in \cite{est2}. The regret for not using the optimal regularized LS estimator is defined by
\begin{equation}\label{Req:LSreg0}
  \R(\vx;\dH,\dy) \defi \left\{ {\norm{\ty - \tH \vx}}^2 + \mu \norm{\vx}^2 \right\}-  \min_{\vw \in \C^n} \left\{ \norm{\ty - \tH \vw}^2 + \mu \norm{\vw}^2 \right\},
\end{equation}
where $\mu > 0$ is the regularization parameter. We emphasize that there are different approaches to choose $\mu$, however, for the focus of this paper, we assume that it is already set before the optimization so that these methods can be readily incorporated in our framework. Hence, we solve the regularized LS estimation problem for an arbitrary $\mu > 0$ and note that we have already covered the $\mu = 0$ case in Section \ref{sec:urls}.

Similar to the previous case, we denote the estimation error of the optimal LS estimator for some estimated data matrix $\vH$ and output vector $\vy$ by
\begin{align}
  f(\vH, \vy) & \defi \min_{\vw \in \C^n}{\norm{\vy - \vH \vw}}^2 + \mu \norm{\vw}^2 \nn\\
              & = \norm{\vP^{-1} \, \vy}^2 \nn\\
              & = \vy^H \vP^{-1} \, \vy, \nn
\end{align}
where $\vP \defi \vI + \mu^{-1}\vH\vH^H$. Considering the first order Taylor series expansion based on Wirtinger calculus \cite{graham} for $f(\tH, \ty)$ around $\tH = \vH$ and $\ty = \vy$
\begin{align}
f(\tH, \ty) & \approx \kappa + 2\operatorname{Re} \left\{ \tr\left( \nabla f(\tH, \ty)\big\vert_{\tH=\vH, \ty = \vy}^H \big[\dH \;\; \dy\big] \right) \right\}, \nn\\
            & = \kappa + \vd^H \dvh + \dvh^H \vd + \vb^H \dy + \dy^H \vb, \nn
\end{align}
where $\vd \defi \col(\vD^H)$, $\dvh \defi \col(\dH)$,
\begin{align}
  \vD & \defi \dfrac{\partial f(\tH, \ty)}{\partial \tH}\bigg\vert_{\tH=\vH, \ty=\vy} \nn\\
      & = - \vP^{-1} \vy \vy^H \vP^{-1} \vH,
\end{align}
and
\begin{align}
  \vb & \defi \dfrac{\partial f(\tH, \ty)}{\partial \ty}\bigg\vert_{\tH=\vH, \ty=\vy} \nn\\
      & = \vP^{-1} \vy, \nn
\end{align}
where the last line follows since $\vP$ is symmetric. Hence we can approximate the regret in \eqref{Req:LSreg0} as follows
\begin{equation}\label{eq:regret_final_2}
  \R(\vx; \dH, \dy) \approx {\norm{\ty - \tH \vx}}^2 + \mu \norm{\vx}^2 - (\kappa + \vd^H \dvh + \dvh^H \vd + \vb^H \dy + \dy^H \vb),
\end{equation}
similar to \eqref{eq:regret_final}. In the following theorem, we illustrate how the optimization problem in \eqref{eq:regret_final_2} can be put in an SDP form.

\begin{thm}\label{thm2}
Let $\vH \in \C^{m\times n}$ and $\vy \in \C^{m}$ be the estimates of the data matrix and the output vector, respectively, both having deterministic additive perturbations $\dH \leq \delta_H$ and $\dy \leq \delta_Y$, respectively, i.e., $\tH=\vH+\dH$ and $\ty=\vy+\dy$, where $\tH$ is the full rank data matrix, $\ty$ is the output vector, and $m \geq n$. Then the problem
\begin{equation}\label{eq:prudefthm2}
  \min_{\vx \in \C^n} \max_{\norm{\dH} \leq \delta_H,\norm{\dy} \leq \delta_Y} \R(\vx; \dH,\dy),
\end{equation}
where $\R(\vx; \dH,\dy)$ is defined as in \eqref{eq:regret_final_2}, is equivalent to solving the following SDP problem
\begin{gather}
  \min \gamma \nn\\
  \mbox{subject to} \nn\\
  \tau_1 \geq 0, \tau_2 \geq 0 \text{, and} \nn\\
  \begin{bmatrix}
  \gamma + \kappa - \tau_1 - \tau_2 & (\vy-\vH\vx)^H & \vx^H   & \delta_Y \vb^H  & \delta_H\vd^H \\
  \vy-\vH\vx                        & \vI            & \vec{0} & -\delta_Y\vI    & \delta_H\vX \\
  \vx                               & \vec{0}        & \mu \vI & \vec{0}         & \vec{0} \\
  \delta_Y \vb                      & -\delta_Y\vI   & \vec{0} & \tau_1 \vI      & \vec{0} \\
  \delta_H\vd                       & \delta_H\vX^H  & \vec{0} & \vec{0}         & \tau_2 \vI
  \end{bmatrix} \hspace{-0.1cm} \geq \hspace{-0.05cm} 0.
\end{gather}
\end{thm}

\begin{pot2}
The proof of Theorem 2 follows similar lines to the proof of Theorem 1, hence is omitted here. \hfill $\square$
\end{pot2}

\begin{rmk}
Under the framework introduced in this section, one can straightforwardly obtain the corollaries similar to Corollaries 1 and 2 by considering cases in which the uncertainty is either only on the data matrix or only on the output vector, i.e., $\delta_Y = 0$ and $\delta_H = 0$ cases, respectively. The derivations follow similar lines to Corollaries 1, 2 and Theorem 2, hence is omitted. However, similar results can be readily derived from the result in Theorem 2 with suitable changes in the SDP formulations.
\end{rmk}

\section{Structured Robust Least Squares Estimation}\label{sec:srls}
There are various communication systems where the data matrix and the perturbation on it have a special structure such as Toeplitz, Hankel, or Vandermonde \cite{est1, tls}. Incorporating this prior knowledge into the estimation framework could improve the performance of the regret based minimax LS estimation approach \cite{est1, tls}. Hence, in this section, we investigate a special case of the problem in \eqref{eq:prudef1}, where the associated perturbations for the data matrix $\vH$ and the output vector $\vy$ have special structures. The structure on the perturbations is defined as follows
\begin{equation}\label{eq:dH}
  \dH = \sum_{i=1}^p\alpha_i\vH_i,
\end{equation}
and
\begin{equation}\label{eq:dy}
  \dy = \sum_{i=1}^p\beta_i\vy_i,
\end{equation}
where $\vH_i \in \C^{m \times n}$, $\vy_i \in \C^m$, and $p$ are known but $\alpha_i,\beta_i \in \C$, $i=1,\dots,p$, are unknown. However, the bounds on the norm of $\va \defi [\alpha_1,\dots,\alpha_p]^H$ and $\vbet \defi [\beta_1,\dots,\beta_p]^H$ are provided as $\norm{\va} \leq \delta_{\alpha}$ and $\norm{\vbet} \leq \delta_{\beta}$, where $\delta_{\alpha}, \delta_{\beta} \geq 0$. We emphasize that this formulation can represent a wide range of constraints on the structure of perturbations of the data matrix and the output vector such as Toeplitz and Hankel \cite{est1, sayedbook}. Our aim is to solve the following optimization problem
\[
\min_{\vx \in \C^n} \max_{\norm{\va} \leq \delta_{\alpha}, \norm{\vbet} \leq \delta_{\beta}} \R(\vx; \dH, \dy),
\]
where
\begin{gather}
  \R(\vx; \dH, \dy) \defi \norm{\ty - \tH \vx}^2 - \min_{\vw \in \C^n} \norm{\ty - \tH \vw}^2, \label{eq:regret_first_3}\\
  \tH \defi \vH + \dH = \vH + \sum_{i=1}^p \alpha_i \vH_i, \label{eq:tH}\\
  \ty \defi \vy + \dy = \vy + \sum_{i=1}^p \beta_i \vy_i. \label{eq:ty}
\end{gather}

After following similar lines to Section \ref{sec:urls}, and introducing the first order Taylor approximation to $f\left(\tH,\ty\right)$ around $\va = \vec{0}$ and $\vbet = \vec{0}$, we obtain
\begin{equation}\label{eq:taylor3}
  f\left(\tH,\ty\right) \approx \kappa + 2\operatorname{Re} \left\{ \tr\left( \nabla f(\tH, \ty)\big\vert_{\va=\vec{0}, \vbet=\vec{0}}^H \left[\va \;\; \vbet\right] \right) \right\},
\end{equation}
where $f\left(\tH,\ty\right) = \ty^H \tP \ty$ and $\tP = \vI - \tH \tH^+$. We next introduce the following lemma to calculate the first order Taylor approximation in \eqref{eq:taylor3} in a closed form.

\begin{lem}\label{lem2}
Let $\tH =\vH+\dH$ be a full rank matrix and $\ty=\vy+\dy$, where $\tH \in \C^{m \times n}$, $\ty \in \C^m$, $\dH$ and $\dy$ are defined as in \eqref{eq:dH} and \eqref{eq:dy}, respectively. Then denoting $f\left(\tH, \ty \right) \triangleq \ty^H\vP\ty$, where $\tP \triangleq \vI-\tH\tH^+$, we have
\begin{equation}
  \dfrac{\partial f(\tH, \ty)}{\partial \va} \bigg\vert_{\va = \vec{0}, \vbet = \vec{0}} = \left[ - \vy^H \vP^H \vH_1 \vH^+ \vy, \dots,  - \vy^H \vP^H \vH_p \vH^+ \vy \right]^H,
\end{equation}
and
\begin{equation}
  \dfrac{\partial f(\tH, \ty)}{\partial \vbet} \bigg\vert_{\va = \vec{0}, \vbet = \vec{0}} = \left[ \vy^H \vP \vy_1, \dots,  \vy^H \vP \vy_p \right]^H,
\end{equation}
where $\vP \triangleq \vI-\vH\vH^+$.
\end{lem}

\begin{pol2}
Note that the derivative of $f\left(\tH, \ty \right)$ is taken with respect to $[\va \;\; \vbet]$, hence we can use the Chain Rule to calculate the derivatives by using the results we have obtained in Lemma 1.

First, we consider the derivative of $f\left(\tH, \ty \right)$ with respect to $\alpha_i$, $i = 1,\dots,p$, i.e.,
\begin{align}
  d_i & \defi \dfrac{\partial f(\tH, \ty)}{\partial \alpha_i} \bigg\vert_{\va = \vec{0}, \vbet = \vec{0}} \nn\\
      & = \tr \left( \left( \dfrac{\partial f(\tH, \ty)}{\partial \tH} \right)^H  \dfrac{\partial \tH}{\partial \alpha_i} \bigg\vert_{\va = \vec{0}, \vbet = \vec{0}} \right) \nn\\
      & = \tr \left( - \vH^+ \vy \vy^H \vP^H \vH_i \right) \nn\\
      & = - \vy^H \vP^H \vH_i \vH^+ \vy, \nn
\end{align}
where the last line follows from the cyclic property of the trace operator.

%From these individual derivatives, we construct the resulting derivative vector
%\begin{align}
%  \vd & \defi \dfrac{\partial f(\tH, \ty)}{\partial \va} \bigg\vert_{\va = \vec{0}, \vbet = \vec{0}} \nn\\
%      & = [d_1,\dots,d_p]^H. \nn
%\end{align}

Similarly, we next consider the derivative of $f\left(\tH, \ty \right)$ with respect to $\beta_i$, $i = 1,\dots,p$, i.e.,
\begin{align}
  b_i & \defi \dfrac{\partial f(\tH, \ty)}{\partial \beta_i} \bigg\vert_{\va = \vec{0}, \vbet = \vec{0}} \nn\\
      & = \tr \left( \left( \dfrac{\partial f(\tH, \ty)}{\partial \ty} \right)^H  \dfrac{\partial \ty}{\partial \beta_i} \bigg\vert_{\va = \vec{0}, \vbet = \vec{0}} \right) \nn\\
      & = \vy^H \vP \vy_i. \nn
\end{align}
This concludes the proof of Lemma 2. \hfill $\square$
\end{pol2}

%From these individual derivatives, we construct the resulting derivative vector
%\begin{align}
%  \vb & \defi \dfrac{\partial f(\tH, \ty)}{\partial \vbet} \bigg\vert_{\va = \vec{0}, \vbet = \vec{0}} \nn\\
%      & = [b_1,\dots,b_p]^H. \nn
%\end{align}

Now turning our attention back to \eqref{eq:taylor3}, we denote
\[
\vd \defi \dfrac{\partial f\left(\tH, \ty \right)}{\partial \va} \bigg\vert_{\va = \vec{0}, \vbet = \vec{0}},
\]
and
\[
\vb \defi \dfrac{\partial f\left(\tH, \ty \right)}{\partial \vbet} \bigg\vert_{\va = \vec{0}, \vbet = \vec{0}},
\]
where we emphasize that the closed form definitions of $\vd$ and $\vb$ can be obtained from Lemma 2. We then approximate \eqref{eq:taylor3} and obtain the first order Taylor approximation as follows
\[
f\left(\tH, \ty \right) \approx \kappa + \vd^H\va + \va^H\vd + \vb^H\vbet + \vbet^H\vb.
\]
Therefore, we can approximate the regret in \eqref{eq:regret_first_3} as follows
\begin{equation}\label{eq:regret_final_3}
  \R(\vx; \dH,\dy) \approx {\norm{\ty - \tH \vx}}^2 - \left( \kappa + \vd^H\va + \va^H\vd + \vb^H\vbet + \vbet^H\vb \right).
\end{equation}

In the following theorem, we illustrate how the optimization problem in \eqref{eq:regret_final_3} can
be put in an SDP form.

\begin{thm}\label{thm3}
Let $\vH, \vH_1, \dots, \vH_p \in \C^{m\times n}$, $\vy, \vy_1,\dots, \vy_p \in \C^{m}$, $\delta_{H}, \delta_{Y} \geq 0$, $m\geq n$, where $\tH$ is the full rank data matrix defined as in \eqref{eq:tH}, $\ty$ is the output vector defined as in \eqref{eq:ty}, with the corresponding estimates $\vH$ and $\vy$, respectively. Then the problem
\begin{equation}\label{eq:prudefthm3}
  \min_{\vx \in \C^n} \max_{\norm{\va} \leq \delta_{\alpha},\norm{\vbet} \leq \delta_{\beta}} \R(\vx; \dH,\dy),
\end{equation}
where $\R(\vx; \dH,\dy)$ is defined as in \eqref{eq:regret_final_3}, is equivalent to solving the following SDP problem
\begin{gather}\label{eq:sdp3}
  \min \gamma \nn\\
  \mbox{subject to} \nn\\
  \tau_1 \geq 0, \tau_2 \geq 0 \text{, and} \nn\\
  \begin{bmatrix}
  \gamma + \kappa - \tau_1 - \tau_2  & (\vy-\vH\vx)^H           & \delta_{\alpha} \vd^H  & \delta_{\beta} \vb^H \\
  \vy-\vH\vx                         & \vI                      & -\delta_{\alpha} \vG   & \delta_{\beta} \vQ \\
  \delta_{\alpha} \vd                & -\delta_{\alpha} \vG^H   & \tau_1 \vI             & \vec{0} \\
  \delta_{\beta} \vb                 & \delta_{\beta} \vQ^H     & \vec{0}                & \tau_2 \vI
  \end{bmatrix} \geq 0,
\end{gather}
where $\vG \defi [\vH_1\vx,\dots,\vH_p\vx]$ and $\vQ \defi [\vy_1,\dots,\vy_p]$.
\end{thm}

\begin{pot3}
The proof of Theorem 2 follows similar lines to the proof of Theorem 1, hence is omitted here. \hfill $\square$
\end{pot3}

\begin{rmk}
Under the framework introduced in this section, one can straightforwardly obtain the corollaries similar to Corollaries 1 and 2 by considering cases in which the uncertainty is either only on the data matrix or only on the output vector, i.e., $\delta_{\beta} = 0$ and $\delta_{\alpha} = 0$ cases, respectively. The derivations follow similar lines to Corollaries 1, 2 and Theorem 3, hence is omitted. However, similar results can be readily derived from the result in Theorem 3 with suitable changes in the SDP formulations.
\end{rmk}

\begin{rmk}
The proofs of Theorem 2 and Theorem 3 follow from the results of Theorem 1, which relies on the lossless {\em S}-procedure. However, {\em S}-procedure is lossless with two constraints when the corresponding two quadratic (Hermitian) forms on the complex linear space \cite{s1}. However, classical {\em S}-procedure for quadratic forms is, in general, lossy with two constraints in the real case \cite{s2}. Hence, Theorem 2 and Theorem 3 cannot be extended for real linear space. On the other hand, under the frameworks described in Remark 3 and Remark 4, one can safely extend the same conclusions for the real case also, since {\em S}-procedure is lossless for quadratic forms with one constraint both in complex and real spaces \cite{huang1,huang2}.
\end{rmk}

\section{Simulations}\label{sec:numer}
We provide numerical examples in different scenarios in order to illustrate the merits of the proposed algorithms. In the first set of the experiments, we randomly generate a data matrix of size $m \times n$, and an output vector of size $m\times 1$, which are normalized to have unit norms. Then, we generate $1000$ random perturbations $\dH$, $\dy$, where $\norm{\dH}\leq \delta_H$, $\norm{\dy}\leq \delta_Y$, $m=5$, $n=3$, and $\delta_H=\delta_Y=1.2$. Here, we label the algorithm in Theorem 1 as ``rgrt-LS'', the robust LS algorithm of \cite{Ghaoui97} as ``rbst-LS'', the total LS algorithm \cite{Ghaoui97} as ``TLS'', and finally the LS algorithm tuned to the estimates of the data matrix and the output vector as ``LS'', where we directly use $\hat{\vx} = \vH^+ \vy$.

\begin{figure}[t]
  \centering
  \includegraphics[width=0.75\textwidth]{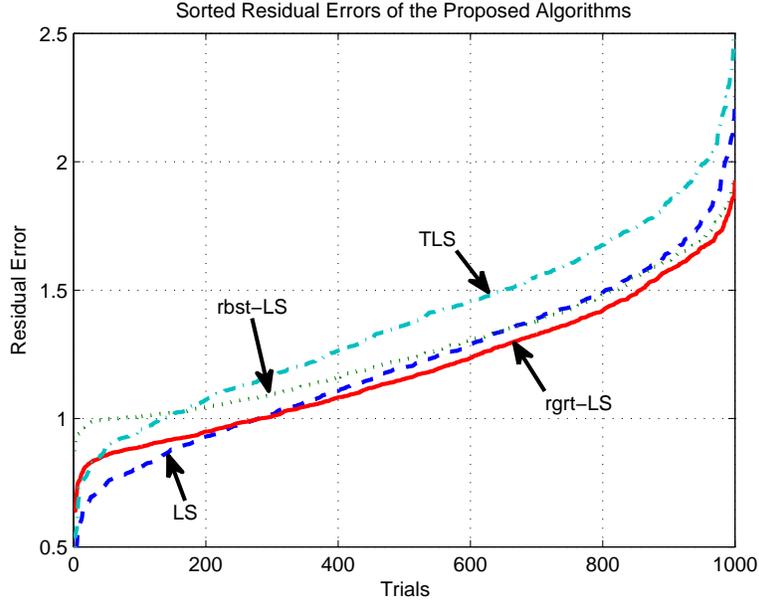}\\
  \caption{Sorted residual errors for the rgrt-LS, rbst-LS, LS, and TLS estimators over $1000$ trials when $\delta_H = \delta_Y = 1.2$, $m=5$, and $n=3$.}\label{fig:ls_ustr}
\end{figure}

For each algorithm and for each random perturbation, we find the corresponding $\hat{\vx}$ and calculate the error $\norm{ \tH \hat{\vx} - \ty}^2$. After we calculate the errors for each algorithm and for all random perturbations, we plot the corresponding sorted errors in ascending order in Fig. 1 for 1000 perturbations. Since the rbst-LS algorithm optimizes the worst-case residual error with respect to worst possible disturbance, it usually yields the smaller worst-case residual error among all algorithms for these simulations. On the other hand, since the LS algorithm directly uses the estimates, it usually yields the smaller residual error when the perturbations on the data matrix and the output vector are significantly small.

These results can be observed in Fig. 1, where in one extreme, the largest residual errors are observed as $2.9762$ for the TLS estimator, $2.2557$ for the LS estimator, $1.9275$ for the rbst-LS estimator, and $1.9325$ for the rgrt-LS estimator. In the other extreme, i.e., when there is almost no perturbation, the smallest estimation errors are observed as $0.3035$ for the LS estimator, $0.4036$ for the TLS estimator, $0.8727$ for the rbst-LS estimator, and $0.6387$ for the rgrt-LS estimator. While the LS estimator can be preferable when there is relatively smaller perturbations and the rbst-LS estimator can be preferable when there is significantly higher perturbations, the introduced algorithm provides a tradeoff between these algorithms and achieve a significantly smaller average error performance. The average residual error of the rgrt-LS estimator is observed as $1.1928$, whereas this value is $1.2180$ for the LS estimator, $1.2708$ for the rbst-LS estimator, and $1.3826$ for the TLS estimator. Hence, the rgrt-LS estimator is not only robust but also efficient in terms of the average error performance compared to its well-known alternatives.

\begin{figure}[t]
  \centering
  \includegraphics[width=0.75\textwidth]{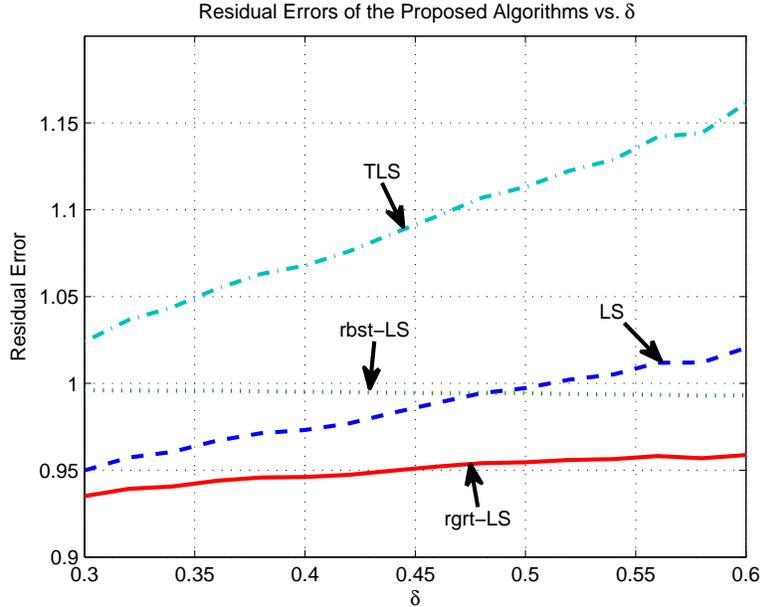}\\
  \caption{Averaged residual errors for the rgrt-LS, rbst-LS, LS, and TLS estimators over $1000$ trials when $\delta_H = \delta_Y \in [0.3, 0.6]$, $m=5$, and $n=3$.}\label{fig:ls_ustr_MSE_vs_rho}
\end{figure}

For the second experiment, we generate $1000$ random perturbations $\dH$, $\dy$, where $\norm{\dH}\leq \delta_H$, $\norm{\dy}\leq \delta_Y$, $m=5$, $n=3$ for different perturbation bounds and compute the averaged error over $1000$ trials for the rgrt-LS, the LS, the rbst-LS, and the TLS algorithms. In Fig. 2, we present the averaged residual errors for these algorithms for different values of perturbation bounds, i.e., $\delta_Y=\delta_H = \delta \in [0.3, 0.6]$. We observe that the proposed rgrt-LS algorithm has the best average residual error performance over different perturbation bounds compared to the LS, the rbst-LS and the TLS algorithms.

\begin{figure}[t]
  \centering
  \includegraphics[width=0.75\textwidth]{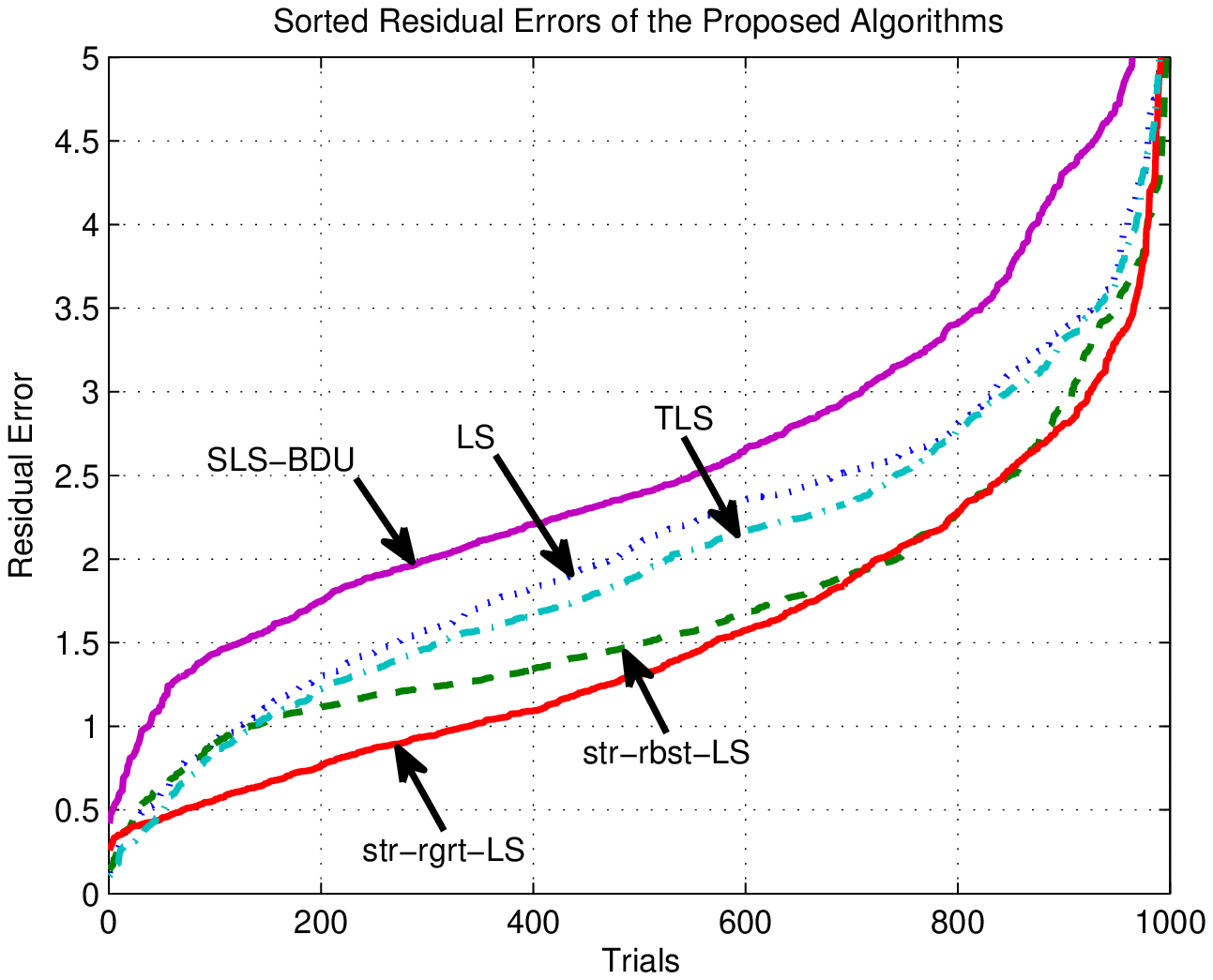}\\
  \caption{Sorted residual errors for the str-rgrt-LS, str-rbst-LS, SLS-BDU, LS, and TLS estimators over $1000$ trials when $\delta_H = \delta_Y = 2$, $m=5$, and $n=3$.}\label{fig:sysid_str}
\end{figure}

As can be observed from Fig. 2, as the perturbation bounds increase, the performances of the LS and the TLS estimators significantly deteriorate, whereas the performance of the rbst-LS estimator almost does not change. The residual error of the rgrt-LS estimator, on the other hand, slightly increases as the perturbation bounds increase, yet the robustness of this estimator can be observed in Fig. 2. Furthermore, the rgrt-LS estimator significantly outperforms its competitors in terms of the average error performance by introducing a transition between the best-case performance of the LS estimator and the worst-case performance of the rbst-LS estimator.

In the next experiment, we examine a system identification problem \cite{pilanci10}, which can be formulated as $\vH_0 \vx=\vy_0$, where $\vH=\vH_0+\vW$ is the observed noisy Toeplitz matrix and $\vy=\vy_0+\vw$ is the observed noisy output vector. Here, the convolution matrix $\vH$ (which is Toeplitz) constructed from $\vh$ which is selected as a random sequence of $\pm 1$'s. For a randomly generated filter $\vh$ of length $3$, we generate $1000$ random structured perturbations for $\vH_0$ and $\vy_0$, where $\norm{\va}\leq 2 \norm{\vH_0}$, and plotted the sorted estimation errors in ascending order in Fig. 3.

\begin{figure}[t]
  \centering
  \includegraphics[width=0.75\textwidth]{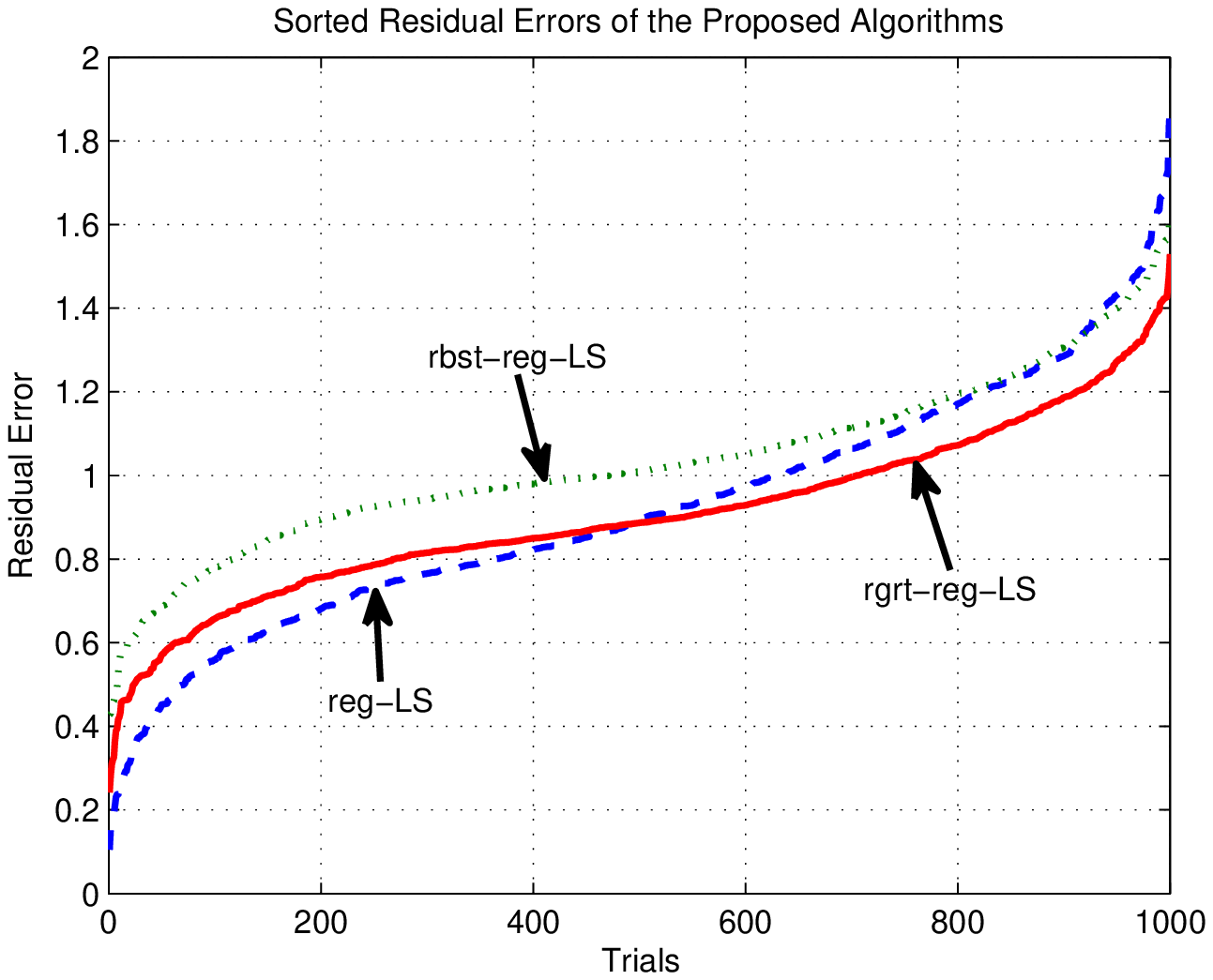}\\
  \caption{Sorted residual errors for rgrt-reg-LS, rbst-reg-LS, and LS estimators over $1000$ trials when $\delta_H = \delta_Y = 0.65$, $\mu = 0.5$, $m=3$, and $n=2$.}\label{fig:sysid_ustr_rls}
\end{figure}

The average residual errors, on the other hand, are observed as $1.5549$ for the structured regret LS estimator ``str-rgrt-LS" of Remark 4, $1.7141$ for the structured robust LS algorithm ``str-rbst-LS'', $2.0061$ for the TLS estimator, $2.1205$ for the LS estimator, and $2.6319$ for the structured least squares bounded data uncertainties estimator, labeled as ``SLS-BDU'' and presented in \cite{pilanci10}. Therefore, we observe that the str-rgrt-LS algorithm yields the smallest average residual error among its competitors. In addition, we observe that the str-rgrt-LS estimator has a smaller residual error in most of the trials compared to its well-known alternatives, owing to its novel regret formulation.

Finally, in Fig. 4, we provide errors sorted in ascending order for the algorithm in Theorem 2 as ``rgrt-reg-LS'', for the robust regularized LS algorithm in \cite{sayed02} as ``rbst-reg-LS'' and finally for the regularized LS algorithm as ``reg-LS'' \cite{Kailath:book}, where the experiment setup is the same as in the first experiment except the perturbation bounds are set to $0.65$ and the regularization parameter is chosen as $\mu=0.5$. In Fig. 4, we observe that the robustness and the performance tradeoff (between the rbst-reg-LS and the reg-LS algorithms) of the introduced rgrt-reg-LS algorithm.

When there is small perturbations on the data matrix and the output vector, i.e., in the best-case scenario, the residual error of the reg-LS estimator is $0.1045$, whereas it is $0.2416$ for the rgrt-reg-LS estimator and $0.4282$ for the rbst-reg-LS estimator. As can be observed from Fig. 4, for higher perturbations, the performance of the reg-LS estimator significantly deteriorates, whereas the rgrt-reg-LS and rbst-reg-LS algorithms provide a robust performance. On the other hand, the rgrt-reg-LS estimator significantly outperforms the rbst-reg-LS estimator in terms of the average error performance and achieves even a more desirable error performance compared to the reg-LS estimator. The average residual errors are calculated as $0.9059$ for the rgrt-reg-LS estimator, $0.9177$ for the reg-LS estimator, and $1.0316$ for the rbst-reg-LS estimator.

\section{Conclusion}\label{sec:conc}
In this paper, we introduce a robust approach to LS estimation problems under bounded data uncertainties based on a novel regret formulation. We study the robust LS estimation problems in the presence of unstructured and structured perturbations under residual and regularized residual error criteria. In all cases, the data vectors that minimize the worst-case regrets are found by solving certain SDP problems. In our simulations, we observed that the proposed estimation methods provide an efficient tradeoff between the performance and robustness, better than the best available alternatives in different signal processing applications.

\appendix
\section{Proof of Theorem 1}\label{app:pot1}
Before we introduce the proof of Theorem 1, we need the following proposition that follows Proposition 2 of \cite{yonina1}.

\begin{prop}\label{prop1}
Given matrices $\vP_1$, $\vQ_1$, $\vP_2$, $\vQ_2$, $\vN$, where $\vN$ is a Hermitian matrix, i.e., $\vN = \vN^H$,
\[
\vN \geq \vP_1^H \vZ_1 \vQ_1 + \vQ_1^H \vZ_1^H \vP_1 + \vP_2^H \vZ_2 \vQ_2 + \vQ_2^H \vZ_2^H \vP_2,
\]
$\forall \: \vZ_1, \vZ_2 : \norm{\vZ_1} \leq \delta_1, \norm{\vZ_2} \leq \delta_2$, if and only if there exist $\tau_1, \: \tau_2 \geq 0$ such that
\begin{equation}\label{eq:prop}
    \begin{bmatrix}
    \vN - \tau_1 \vQ_1^H \vQ_1 - \tau_2 \vQ_2^H \vQ_2 &  -\delta_1 \vP_1^H & -\delta_2 \vP_2^H \\
    -\delta_1 \vP_1                                   & \tau_1\vI          & \vec{0} \\
    -\delta_2 \vP_2                                   & \vec{0}            & \tau_2 \vI\\
    \end{bmatrix} \geq 0.
\end{equation}
\end{prop}

\begin{pop1}
Following similar lines to \cite{yonina1}, we first note that
\[
\vN \geq \vP_1^H \vZ_1 \vQ_1 + \vQ_1^H \vZ_1^H \vP_1 + \vP_2^H \vZ_2 \vQ_2 + \vQ_2^H \vZ_2^H \vP_2,
\]
$\forall \: \vZ_1, \vZ_2 : \norm{\vZ_1} \leq \delta_1, \norm{\vZ_2} \leq \delta_2$, if and only if for every $\vu$ we have
\begin{align}\label{cauch1}
    \vu^H \vN \vu & \geq \max_{\norm{\vZ_1} \leq \delta_1, \: \norm{\vZ_2} \leq \delta_2} \left\{ \vu^H \vP_1^H \vZ_1 \vQ_1 \vu + \hspace{-0.05cm} \vu^H \vQ_1^H \vZ_1^H \vP_1 \vu + \vu^H \vP_2^H \vZ_2 \vQ_2 \vu + \vu^H \vQ_2^H \vZ_2^H \vP_2 \vu \right\} \nn\\
    & = 2 \delta_1 \norm{\vP_1 \vu} \norm{\vQ_1 \vu} + 2 \delta_2 \norm{\vP_2 \vu} \norm{\vQ_2 \vu},
\end{align}
where the last line follows from the Cauchy-Schwartz inequality by choosing
\[
\vZ_1 = \frac{\delta_1 \vP_1 \vu \vu^H \vQ_1^H }{ \norm{\vP_1 \vu} \norm{\vQ_1 \vu} },
\]
and
\[
\vZ_2 = \frac{\delta_2 \vP_2 \vu \vu^H \vQ_2^H }{ \norm{\vP_2 \vu} \norm{\vQ_2 \vu} }.
\]
Furthermore, from the Cauchy-Schwartz inequality, \eqref{cauch1} can be written as
\begin{equation}\label{eq:cauch2}
  \vu^H \vN \vu - 2 \left( \delta_1 \vy_1^H \vP_1 \vu + \delta_2 \vy_2^H \vP_2 \vu \right) \geq 0,
\end{equation}
$\forall \vu, \vy_1, \vy_2 : \norm{\vy_1} \leq \norm{\vQ_1 \vu}, \norm{\vy_2}\leq \norm{\vQ_2\vu}$. Note that the constraint $\norm{\vy_1}\leq \norm{\vQ_1 \vu}$ is equivalent to
\[
\vu^H \vQ_1^H \vQ_1 \vu - \vy_1^H \vy_1 \geq 0,
\]
and similarly, $\norm{\vy_2}\leq \norm{\vQ_2 \vu}$ is equivalent to
\[
\vu^H \vQ_2^H \vQ_2 \vu - \vy_2^H \vy_2 \geq 0.
\]
Hence, after some algebra we obtain \eqref{eq:cauch2} as follows
\[
\begin{bmatrix} \vu \\ \vy_1 \\ \vy_2 \end{bmatrix}^H
\underbrace{\begin{bmatrix}
\vN              & -\delta_1 \vP_1^H & -\delta_2 \vP_2^H \\
-\delta_1 \vP_1  & \vec{0}           & \vec{0} \\
-\delta_2 \vP_2  & \vec{0}           & \vec{0}
\end{bmatrix}}_{\defi \vF_0}
\underbrace{\begin{bmatrix} \vu \\ \vy_1 \\ \vy_2 \end{bmatrix}}_{\defi \vy}
\geq 0,
\]
$\forall \vy$ such that
\[
\begin{bmatrix} \vu \\ \vy_1 \\ \vy_2 \end{bmatrix}^H
\underbrace{\begin{bmatrix}
\vQ_1^H \vQ_1 & \vec{0} & \vec{0} \\
\vec{0}       & -\vI    & \vec{0} \\
\vec{0}       & \vec{0} & \vec{0}
\end{bmatrix}}_{\defi \vF_1}
\begin{bmatrix} \vu \\ \vy_1 \\ \vy_2 \end{bmatrix}
\geq 0,
\]
and
\[
\begin{bmatrix} \vu \\ \vy_1 \\ \vy_2 \end{bmatrix}^H
\underbrace{\begin{bmatrix}
\vQ_2^H \vQ_2 & \vec{0} & \vec{0} \\
\vec{0}       & \vec{0} & \vec{0} \\
\vec{0}       & \vec{0} & -\vI
\end{bmatrix}}_{\defi \vF_2}
\begin{bmatrix} \vu \\ \vy_1 \\ \vy_2 \end{bmatrix}
\geq 0.
\]
Then applying {\em S}-procedure \cite{boyd}, we have
\begin{eqnarray}\label{eq:ref1}
  && \vy^H \vF_0 \vy \geq 0, \nn\\
  && \forall \vy : \vy^H \vF_1 \vy \geq 0, \vy^H \vF_2 \vy \geq 0, \nn\\
  && \text{where }\exists \: \vy_0 : \vy_0^H \vF_1 \vy_0\ > 0, \vy_0^H \vF_2 \vy_0> 0.
\end{eqnarray}
Note that due to the structures of $\vF_1$ and $\vF_2$, the regularity conditions can be easily verified. Since $\vF_0, \vF_1, \text{ and }\vF_2$ are Hermitian matrices, i.e., $\vF_i = \vF_i^H$, $i=0,1,2$, by Theorem 1.1 in \cite{s1}, \eqref{eq:ref1} is satisfied if and only if $\exists \: \tau_1, \tau_2 \geq 0$ such that
\[
\vF_0-\tau_1\vF_1-\tau_2\vF_2 \geq 0.
\]
That is
\[
\begin{bmatrix}
\vN - \tau_1 \vQ_1^H \vQ_1 - \tau_2 \vQ_2^H \vQ_2   & -\delta_1 \vP_1^H & -\delta_2 \vP_2^H  \\
-\delta_1\vP_1                                      & \tau_1 \vI        & \vec{0}  \\
-\delta_2\vP_2                                      &  \vec{0}          & \tau_2 \vI  \\
\end{bmatrix} \geq 0.
\]
This concludes the proof of Proposition 1. \hfill $\square$
\end{pop1}

Now we consider the minimax problem defined in \eqref{eq:prudefthm1}, and reformulate it as follows
\[
  \min_{\vx \in \C^n} \max_{\norm{\dH} \leq \delta_H,\norm{\dy} \leq \delta_Y} \R(\vx; \dH,\dy) = \min_{\vx,\gamma} \gamma,
\]
subject to % {\norm{\ty - \tH \vx}}^2 - \left( f(\vH, \vy) + \vd^H \dvh + \dvh^H \vd + \vb^H \dy + \dy^H \vb \right)
\begin{equation}\label{eq:here}
  \R(\vx; \dH,\dy) \leq \gamma, \; \forall \dH, \dy : \norm{\dH} \leq \delta_H, \norm{\dy} \leq \delta_Y,
\end{equation}
where
\begin{equation}
  \R(\vx; \dH,\dy) = {\norm{\ty - \tH \vx}}^2 - \left( \kappa + \vd^H \dvh + \dvh^H \vd + \vb^H \dy + \dy^H \vb \right),
\end{equation}
and $\kappa \defi f(\vH, \vy)$. By applying the Schur complement to the constraints in \eqref{eq:here}, we can compactly denote \eqref{eq:here} in the matrix form as follows
\begin{equation}\label{sdpi0}
  \begin{bmatrix}
  \gamma \hspace{-0.1cm} + \hspace{-0.1cm} \kappa \hspace{-0.1cm} + \hspace{-0.1cm} \vd^H \dvh \hspace{-0.1cm} + \hspace{-0.1cm} \dvh^H \vd \hspace{-0.1cm} + \hspace{-0.1cm} \vb^H \dy \hspace{-0.1cm} + \hspace{-0.1cm} \dy^H \vb & \left( \ty - \tH \vx \right)^H \\
  \ty - \tH \vx    & \vI
  \end{bmatrix}  \hspace{-0.1cm}
  \geq \hspace{-0.05cm} 0,
\end{equation}
$\forall \dH, \dy : \norm{\dH} \leq \delta_H, \norm{\dy} \leq \delta_Y$. Rearranging terms in \eqref{sdpi0}, we obtain
\begin{align}\label{sdpi1}
  \begin{bmatrix}
  \gamma + \kappa   & \left( \vy - \vH \vx \right)^H  \\
  \vy - \vH \vx     & \vI
  \end{bmatrix}
  & \geq -
  \begin{bmatrix} \vd^H \\ \vX \end{bmatrix}
  \dvh
  \begin{bmatrix} 1 & \vec{0} \end{bmatrix}
  -
  \begin{bmatrix} 1 \\ \vec{0} \end{bmatrix}
  \dvh^H
  \begin{bmatrix} \vd & \vX^H \end{bmatrix} \nn\\
  & \hspace{1.5cm} -
  \begin{bmatrix} \vb^H \\ -\vI \end{bmatrix}
  \dy
  \begin{bmatrix} 1 & \vec{0} \end{bmatrix}
  -
  \begin{bmatrix} 1 \\ \vec{0} \end{bmatrix}
  \dy^H
  \begin{bmatrix} \vb & -\vI \end{bmatrix},
\end{align}
$\forall \dH, \dy : \norm{\dH} \leq \delta_H, \norm{\dy} \leq \delta_Y$, where we used $\dH \vx = \vX\dvh$, $\dvh = \mathrm{vec}\left(\dH\right)$, and $\vX \defi \vx^H \otimes \vI$. By applying Proposition 1 to \eqref{sdpi1}, it follows that \eqref{eq:prudefthm1} is equivalent to
\begin{gather}
  \min \gamma \nn\\
  \mbox{subject to} \nn\\
  \tau_1 \geq 0, \tau_2 \geq 0 \text{, and} \nn\\
  \begin{bmatrix}
  \gamma + \kappa - \tau_1 - \tau_2   & (\vy - \vH \vx)^H  & \delta_Y \vb^H  & \delta_H \vd^H  \\
  \vy - \vH \vx                       & \vI                & -\delta_Y \vI   & \delta_H \vX  \\
  \delta_Y \vb                        & -\delta_Y \vI      & \tau_1 \vI      & \vec{0}  \\
  \delta_H \vd                        & \delta_H \vX^H     & \vec{0}         & \tau_2 \vI
  \end{bmatrix} \geq 0, \nn
\end{gather}
hence the desired result. Therefore, this concludes the proof of Theorem 1. \hfill $\square$

\bibliographystyle{elsarticle-num}

\end{document}